\documentclass[aps,pre,twocolumn,float]{revtex4}
\usepackage{amsmath,bm,epsfig}
\let \nn  \nonumber
\usepackage{pstricks}
\usepackage{pst-node}
\usepackage[ansinew]{inputenc}
\usepackage{amssymb,amsmath}
\usepackage{multirow}

\def\be{\begin{equation}}       \def\ba{\begin{array}}

\def\ee{\end{equation}}         \def\ea{\end{array}}

\def\bea {\begin{eqnarray}}      \def\eea {\end{eqnarray}}

\def\bean{\begin{eqnarray*}}    \def\eean{\end{eqnarray*}}

\newtheorem{exi}{Example}

\begin{document}

\title{Universal power law for the energy spectrum of breaking Riemann waves}
\author{Dmitry Pelinovsky$^{a,b}$, Efim Pelinovsky$^{b,c,d}$, Elena Kartashova$^{d}$, Tatjana Talipova$^{b,c}$, and Ayrat Giniyatullin$^{b}$}
   \affiliation{$^a$Department of Mathematics, McMaster University, Hamilton, Ontario, Canada}
   \affiliation{$^b$Department of Applied Mathematics,  Nizhny Novgorod State Technical University, Nizhny Novgorod, Russia}
    \affiliation{$^c$Department of Nonlinear Geophysical Processes, Institute of Applied Physics, Nizhny Novgorod, Russia}
     \affiliation{$^d$Institute for Analysis, Johannes Kepler University, Linz, Austria}

\begin{abstract}
The universal power law for the spectrum of one-dimensional breaking Riemann waves
is justified for the simple wave equation. The spectrum of spatial amplitudes at the breaking time
$t = t_b$ has an asymptotic decay of $k^{-4/3}$, with corresponding energy spectrum decaying as $k^{-8/3}$.
This spectrum is formed by the singularity of the form $(x-x_b)^{1/3}$ in the wave shape
at the breaking time. This result remains valid for arbitrary nonlinear wave speed.
In addition, we demonstrate numerically that the universal power law is observed for long time in the range of
small wave numbers if small dissipation or dispersion is accounted in the viscous Burgers or Korteweg-de Vries equations.
\end{abstract}


\maketitle
\section{Introduction}

One-dimensional traveling nonlinear waves in dispersiveless systems are called
Riemann or simple waves; their dynamics is well studied in various physical media,
e.g. acoustic waves in the compressible fluids and gases \cite{RS77,GMS91}, surface
and internal waves in oceans \cite{DZKP06,SR07,ZSTPKP07,ZDKP08,OH11},
tidal and tsunami waves in rivers \cite{TYMF91, Chan11}, ion- and magneto-sound
waves in plasmas \cite{Plasma}, electromagnetic waves in transmission lines \cite{GOF67},
and optical tsunami in fiber optics \cite{W}.
In homogeneous and stationary media, the Riemann waves continuously deform and
transform to the shock waves yielding breaking in a finite time.

In nonlinear acoustics where the wave intensity is not very high,
the nonlinear deformation of the Riemann wave has been studied in many details
since this stage occurs during many wavelengths \cite{Whith74,Pel76,EFP88,OP99}.
Corresponding  nonlinear evolution equation is a so-called simple wave equation
\be\label{e1}
u_t+V(u) u_x = 0,
\ee
where $u$ is a wave function, and $V(u)$ is a characteristic local velocity
of the various points of the wave shape. If $V'(u) > 0$ for all $u$
(that is, if $V$ is invertible), equation (\ref{e1}) can be
written for the local velocity:
\be\label{e19}
V_t + V V_x = 0,
\ee
which is equivalent to the inviscid Burgers equation
\be\label{e2}
v_t + v v_x = 0, \quad v := V(u).
\ee
The Cauchy problem for equation (\ref{e2}) starts with
the initial condition given by a smooth function
that decays to zero at infinity:
\be\label{e3}
v(x,0) = F(x), \quad \lim_{|x| \to \infty} F(x) = 0,
\ee
and results in the implicit solution called a simple or Riemann wave:
\be\label{e4}
v(x,t) = F\left(x - t v(x,t)\right).
\ee
We are interested in the asymptotic behavior of these solutions near the breaking time
$t = t_b$. The appearance of the singularity in the wave shape yields a power law in the
Fourier spectrum of wave turbulence \cite{Kuz04}. It was argued earlier in \cite{GSYa83}
that for simple (Riemann) waves this singularity is of the form $v - v_b \sim (x_b-x)^{1/2}$;
this corresponds to the spectrum of spatial amplitudes decaying at the breaking time
as $k^{-3/2}$, with corresponding energy spectrum decaying as $k^{-3}$.

However, we will show that this assumption is in fact incorrect and
the spectrum of spatial amplitudes decays at the breaking time
as $k^{-4/3}$, with corresponding energy spectrum decaying as $k^{-8/3}$,
because the simple waves develop singularities of the form $v - v_b \sim (x-x_b)^{1/3}$.
This analytical result confirms earlier numerical observations in Ref. \cite{KP13}.

We note that the wave field in the vicinity of the breaking time was studied by Pomeau {\em et al.} \cite{Pomeaux1,Pomeaux2}.
They observed that the wave profile changes as $v \sim (t_b - t)^{1/2}$ near the breaking time;
to show this, they developed Taylor series expansions similar to methods in our paper.
Although the result $v - v_b \sim (x-x_b)^{1/3}$ can be derived from the formulas given in \cite{Pomeaux1,Pomeaux2},
the authors did not point it explicitly and did not study the Fourier spectrum of the breaking Riemann wave.

Let us now illustrate the wave breaking and the main result with examples. First,
it is easy to see that the wave steepness $v_x$ increases on the wave front
where $F'$ is negative:
\be \label{e5}
v_x = \frac{F'(\zeta)}{1+t F'(\zeta)}, \quad \zeta = x - t v(x,t).
\ee
The breaking time is computed explicitly as
\be \label{e6}
t_b = \frac{1}{\max_{x}(-F'(x))} = \frac{-1}{\min_x(F'(x))}.
\ee

\begin{figure}
\begin{center}
\includegraphics[width=8.5cm]{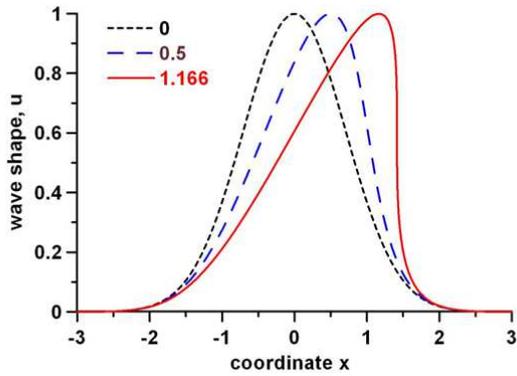}
\caption{\label{f:1} Deformation of the Riemann wave for the case of quadratic
nonlinearity and initial pulse of Gaussian shape $F(x)=\exp{(-x^2)}.$
The wave shape at the moments of time $t=0, 0.5, 1.166$ is shown as small dashed,
long dashed and bold lines correspondingly.}
\end{center}
\end{figure}

This process is illustrated at Fig.\ref{f:1} for an initial pulse of the Gaussian shape
$F(x) =\exp{(-x^2)}$. The shock is formed at the point $x_b = 2^{-1/2}$ and $v_b =\exp{(-1/2)}$
at the moment of time $t_b = 2^{-1/2}\exp{(1/2)}\approx 1.166$.
At the breaking time the wave shape contains singularity on its front (i.e. its
steepness becomes infinite) and the solution (\ref{e4}) is not valid anymore.

\begin{figure}
\begin{center}
\includegraphics[width=9.5cm]{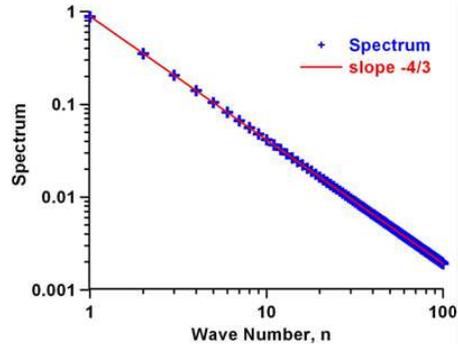}
\caption{\label{f:2} Amplitude spectrum of Riemann waves for the solution (\ref{e7})
at the breaking time $t_b = 1$ is shown by blue crosses.
The power rate $n^{-4/3}$ is shown as the red solid line.  Wave number $n$ (horizontal axes) and
the amplitude spectrum $|A_n|$ (vertical axes) are  shown in logarithmical coordinates.}
\end{center}
\end{figure}

If $f(x) = \sin(x)$, an analytical solution for the Fourier spectrum of the simple waves
(\ref{e4}) is known. Corresponding spectrum is known as the Bessel-Fubini spectrum and is given
in Ref. \cite{Pel76}:
\be \label{e7}
v(x,t)=\sum_{n=1}^{\infty} \frac{2 (-1)^{n-1}}{nt} J_n(n t) \sin{(nx)},
\ee
where $J_n$ is  Bessel function of the first order with integer $n$ and the breaking time is
$t_b = 1$. The amplitude of the Fourier spectrum  at the breaking time reads
\be\label{e8}
|A_n|=\frac{2}{n}J_n(n)
\ee
and is distributed close to the rate of $n^{-4/3}$ (see Fig.\ref{f:2}).

We shall now prove that the power rate of the Fourier amplitude spectrum
at the time of wave breaking is $k^{-4/3}$
for a simple (Riemann) wave supported by an arbitrary smooth initial pulse and
an arbitrary local velocity $V(u)$. Moreover, the same rate remains valid in the
range of small wave numbers if small dissipation or dispersion is added in the framework of the
viscous Burgers or Korteweg--de Vries equations.

\section{Power law of wave breaking}

Using the method of characteristics, we write the inviscid Burgers equation
(\ref{e2}) as the system of two ordinary differential equations
\be \label{e9-10}
\frac{d x}{d t} = v(x,t), \qquad \frac{d v}{d t}=0,
\ee
therefore, each point on the wave shape moves with velocity proportional to
the magnitude of $v$. The solution is now written in the parametric form:
\be  \label{e11-12}
u(t)=F(\zeta), \qquad x(t)=\zeta+tF(\zeta).
\ee


Let $\zeta_b$ be the global minimum of $F'$ (which always exists since
$F$ is smooth and decays to zero at infinity). We assume that the minimum
is not degenerate, hence, $F''(\zeta_b) = 0$ and $F'''(\zeta_b) > 0$.
Let $t_b$ be the time of breaking defined by equation (\ref{e6})
such that $1 + t_b F'(\zeta_b) = 0$. The wave breaks at
the point $x_b = \zeta_b + t_b F(\zeta_b)$ and $v_b = F(\zeta_b)$.
Using the decomposition $\zeta = \zeta_b + \eta$ and expanding
the exact solution (\ref{e11-12}) into Taylor series, we obtain
at the time of breaking:
\bea\label{e15}
x & = & \zeta_b + \eta + t_b F(\zeta_b+\eta) \nn \\
& = & x_b +\frac{1}{6} t_b F'''(\zeta_b) \eta^3 + \mathcal{O}(\eta^4)
\eea
and
\bea \label{e17}
v(x,t_b) &=& F(\zeta_b+\eta) \nn\\
&=& v_b + F'(\zeta_b) \eta + \mathcal{O}(\eta^2).
\eea

Solving  (\ref{e15}) for a unique small real root of $\eta$,
we obtain an explicit relation between $u$ and $x$ at the time of breaking:
\be \label{e16}
\eta \sim (x-x_b)^{1/3}, \quad v - v_b \sim (x-x_b)^{1/3}.
\ee
Therefore, the wave profile changes near the breaking point at the breaking time
as $(x-x_b)^{1/3}$, contrary to the behavior $(x-x_b)^{1/2}$ suggested in \cite{GSYa83}.
The behavior (\ref{e16}) leads to the power spectrum with slope $-4/3$ as it was established
numerically in \cite{KP13}. The energy spectrum in this case has the slope $-8/3$.

If $F'''(\zeta_b) = 0$, then $F^{(4)}(\zeta_b) = 0$ since $F$ is smooth and
$\zeta_b$ is the point of minimum of $F'$. If $F^{(5)}(\zeta_b) \neq 0$ (in which case
$F^{(5)}(\zeta_b) > 0$), then the modification of the previous analysis shows that
the wave profile changes near the breaking point at the breaking time
as $(x-x_b)^{1/5}$, leading to the power spectrum with slope $-6/5$.
We can continue this analysis if $\zeta_b$ is a degenerate minimum of a higher order.

Note that the above analysis holds for a general nonlinear evolution equation
(\ref{e1}) under the assumption that $V'(u) > 0$ (that is, when $V$ is invertible).
In this case, if $F'''(\zeta_b) \neq 0$, the wave field $u(x,t_b)$
of the nonlinear evolution equation (\ref{e1}) changes according to
the behavior (\ref{e16}), or explicitly, as
\bea\label{e20}
u(x,t_b) &\approx & V^{-1}\left( v_b + \alpha (x-x_b)^{1/3} \right) \nn \\
&\approx& u_b + \frac{\alpha}{V'(u_b)} (x-x_b)^{1/3},
\eea
where $V^{-1}$ is the inverse function to $V$, $u_b = V^{-1}(v_b)$, and
$\alpha \neq 0$ is a numerical coefficient. Thus, we conclude that
the above universal behavior extends to a general nonlinear evolution
equation (\ref{e1}) and a general initial data (\ref{e3}) under some
restrictive assumptions that are physically relevant.

\section{Small dissipation and dispersion effects}

The simple wave equation (\ref{e1}) is only valid before the moment of breaking;
the study of the wave field evolution at the later times is usually conducted by
including effects of dissipation or dispersion. Corresponding  terms  added to the
right hand side of (\ref{e1}) produce different types of equations such as
the viscous Burgers equation
\be\label{BE}
u_t+ u u_x  = \nu u_{xx}, \quad \nu > 0
\ee
and the Korteweg-de Vries equation
\be\label{KdV}
u_t+6 u u_x + u_{xxx}= 0.
\ee
In both cases, we consider the initial-value problem
starting with initial data $u(x,0)=S_0\sin{( x)}$.

Taking into account dissipative and dispersive effects will inevitably
change the wave spectrum for large wave numbers. For instance, it is
well-known that shock waves in the viscous Burgers
equation (\ref{BE}) have  spectral density that decays as
$k^{-2}$ for large wave numbers \cite{GMS91}.

\begin{figure}
\includegraphics[width=7.5cm]{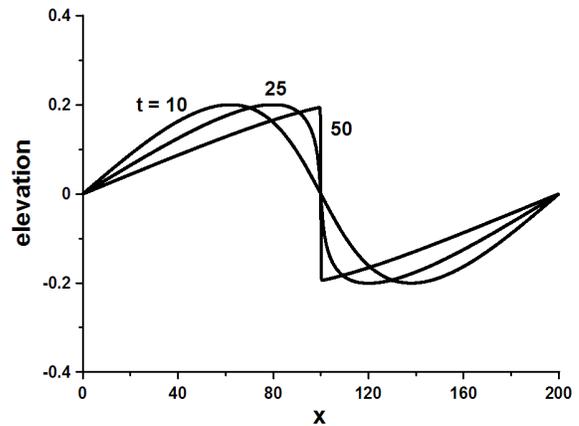}
\caption{\label{f:5} Solutions of the Burgers equation (\ref{BE})
with $\nu=0.1$ and initial data $u(x,0)=S_0\sin{(x)}$ in physical space,
for different values of time $t=10, 20, 30$. }
\end{figure}

\begin{figure}
\includegraphics[width=7.5cm]{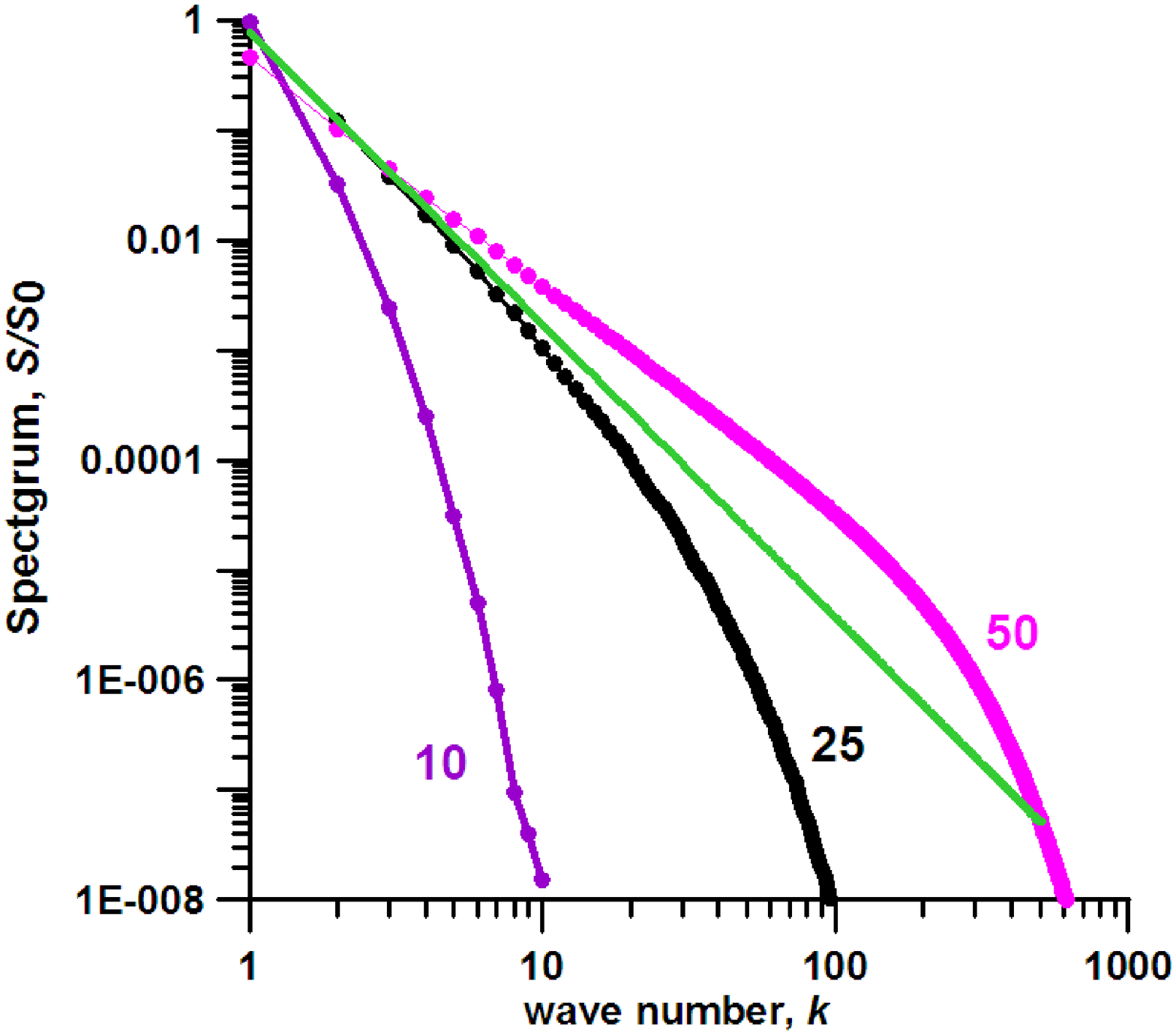}
\caption{\label{f:6} Solutions of the Burgers equation (\ref{BE})
as in Fig.\ref{f:5} but in Fourier space: the normalized energy spectrum $S/S_0$ is shown
versus wave numbers $k$, with axes in logarithmic coordinates. The universal power law $k^{-8/3}$ is
 shown by green solid line.}
\end{figure}

Nevertheless, our numerical simulations of the viscous Burgers equation
(\ref{BE}) demonstrate that
universal power $k^{-8/3}$ of the energy spectrum of breaking Riemann waves is clearly visible
in the range of small wave numbers at least for the evolution times $t \sim t_b \approx 25.5$,
where $t_b$ is the breaking time in the inviscid Burgers equation. For longer times,
dissipative effects become fully developed and drift the energy spectrum away
from the power $k^{-8/3}$. Figures \ref{f:5} and \ref{f:6} illustrate solutions
of the viscous Burgers equation (\ref{BE}) in physical and Fourier space consequently.

\begin{figure}
\includegraphics[width=7.5cm]{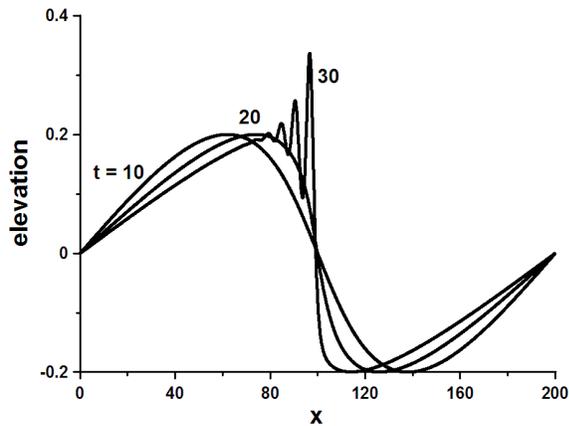}
\caption{\label{f:3} Solutions of the KdV equation (\ref{KdV}) with initial data $u(x,0)=S_0\sin{( x)}$
in physical space, for different values of time $t=10, 20, 30$. }
\end{figure}

\begin{figure}
\includegraphics[width=7.5cm]{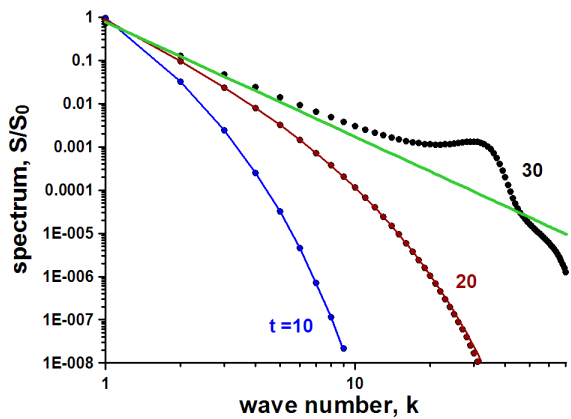}
\caption{\label{f:4} Solutions of the KdV equation (\ref{KdV}) as in Fig.\ref{f:3} but in Fourier space:
the normalized energy spectrum $S/S_0$ is shown versus wave numbers $k$, with axes in logarithmic
coordinates. The universal power law $k^{-8/3}$ is shown by green solid line.}
\end{figure}

Similar results of numerical simulations for the
Korteweg-de Vries equation (\ref{KdV}) are shown in Figures \ref{f:3} and \ref{f:4}.
It is clearly visible that although wave breaking is absent in the KdV equation (\ref{KdV}),
the universal power $k^{-8/3}$ appear in the energy spectrum for small wave numbers for $t \sim t_b \approx 25.5$.



We also mention results of the numerical simulations
of the reduced Ostrovsky equation \cite{Ostr-Eq}
\be \label{Ostr-eq}
( \eta_t + \eta \eta_x )_x = \gamma \eta, \quad \gamma > 0,
\ee
starting with the same initial data $u(x,0)=S_0\sin{( x)}$. A similar effect is observed, namely,
the universal power $k^{-8/3}$ appear in the energy spectrum for large wave numbers
regardless of the rotation parameter $\gamma$ and the initial wave amplitude $S_0$.
The universal behavior is now observed in the range of large wave numbers because the dispersion term in
the reduced Ostrovsky equation (\ref{Ostr-eq}) affects the wave dispersion for small wave numbers.

\section{Summary}

We have justified the universal power law $k^{-8/3}$ in the energy spectrum of one-dimensional breaking Riemann waves
in the context of the simple wave equation (\ref{e1}) with smooth initial data (\ref{e3}). This result remains
valid for arbitrary nonlinear wave speed
provided that the wave speed is an invertible function of the wave amplitude.
In addition, we have demonstrated that the same power law is observed for long times in the range of
small wave numbers in the context of the viscous Burgers (\ref{BE}) and
Korteweg-de Vries (\ref{KdV}) equations. These universal power law also occurs in
other nonlinear evolution equations that reduce to the simple wave equation in
the dissipationless and dispersionless limit.

\vspace{1cm}

\textbf{{Acknowledgments.}}
Part of this work was accomplished during the topical program on ``{\em Mathematics of Oceans}" (Fields Institute, Toronto,
Canada; April-June 2013). The work of DP and AG is supported by the ministry of education
and science of Russian Federation (Project 14.B37.21.0868). The work of EP and TT is supported by
VolkswagenStiftung, RFBR grants (12-05-00472, 13-05-90424).
EK is supported by the Austrian Science Foundation (FWF) under projects	 P22943 and P24671. TT is supported by  RFBR grant 12-05-33070.

\end{document}